\documentclass{aa}

\usepackage{graphicx}
\usepackage[varg]{txfonts}
\usepackage{natbib}
\usepackage[pdftex,colorlinks=true,linkcolor=blue,citecolor=blue,urlcolor=blue]{hyperref}
\usepackage{amsmath}
\usepackage{bbold}
\usepackage{tikz}
\usetikzlibrary{decorations.pathreplacing,math}

\def\earth{\oplus}

\def\expo#1{\mathbf{e}^{#1}}

\def\G{\mathcal{G}}
\def\H{\mathcal{H}}
\def\K{\mathcal{K}}

\def\t{^\mathrm{T}}
\def\mt{^\mathrm{-T}}

\def\id{\mathbb{1}}

\def\L{\mathcal{L}}
\def\O{\mathcal{O}}
\def\mean#1{\overline{#1}}

\def\spleafURL{\url{https://gitlab.unige.ch/jean-baptiste.delisle/spleaf}}
\def\leaf{\textsc{leaf}}
\def\spleaf{\textsc{s+leaf}}
\def\celerite{\textit{celerite}}

\DeclareMathOperator{\diag}{diag}
\DeclareMathOperator{\tril}{tril}
\DeclareMathOperator{\triu}{triu}

\begin{document}

\title{Efficient modeling of correlated noise}
\subtitle{II. A flexible noise model with fast and scalable methods}
\author{J.-B. Delisle\inst{1}
  \and N. Hara\inst{1,}\thanks{NCCR CHEOPS fellow}
  \and D. Ségransan\inst{1}
}
\institute{Département d'astronomie, Université de Genève,
  51 chemin des Maillettes, 1290 Versoix, Suisse\\
  \email{jean-baptiste.delisle@unige.ch}
}

\date{\today}

\abstract{
  Correlated noise affects most astronomical datasets
  and to neglect accounting for it can lead to spurious signal detections,
  especially in low signal-to-noise conditions,
  which is often the context in which new discoveries are pursued.
  For instance, in the realm of exoplanet detection with radial velocity time series,
  stellar variability can induce false detections.
  However, a white noise approximation is often used because
  accounting for correlated noise when analyzing data
  implies a more complex analysis.
  Moreover, the computational cost can be prohibitive
  as it typically scales as the cube of the dataset size.

  For some restricted classes of correlated noise models,
  there are specific algorithms that can be used to help bring down the computational cost.
  This improvement in speed is particularly useful in the context of Gaussian process regression,
  however, it comes at the expense of the generality of the noise model.

  In this article, we present the \spleaf{} noise model, which allows us to
  account for a large class of correlated noises
  with a linear scaling of the computational cost with respect to the size of the dataset.
  The \spleaf{} model includes, in particular, mixtures of quasiperiodic kernels and calibration noise.
  This efficient modeling is made possible by a sparse representation of the
  covariance matrix of the noise and the use of dedicated algorithms for matrix inversion,
  solving, determinant computation, etc.

  We applied the \spleaf{} model to reanalyze the HARPS radial velocity time series of
  the recently published planetary system \object{HD~136352}.
  We illustrate the flexibility of the \spleaf{} model in handling various sources of
  noise.
  We demonstrate the importance of taking correlated noise into account, and
  especially calibration noise, to correctly assess the significance of detected signals.

  We provide an open-source reference implementation of the \spleaf{} model,
  the spleaf package (C library with python wrappers),
  available at \spleafURL.
}


\maketitle

\section{Introduction}
\label{sec:intro}

Astronomical datasets, like most datasets, are contaminated by various sources of noise,
such as photon noise, the intrinsic variability of the object of interest,
contamination by the Earth's atmosphere, instrumental noise, etc.
While the photon noise is purely white (i.e., uncorrelated),
most of the other sources of noise have temporal or spatial correlations.
When neglected, these correlations can lead to spurious signal detections.

In the context of exoplanet detection with radial velocity time series,
stellar variability could induce signals
that mimic planetary signatures \citep[e.g.,][]{queloz_planet_2001}.
The mitigation of stellar variability has become a major subject in planet search studies
and is now routinely achieved by modeling it as correlated Gaussian noise.
Adopting such models significantly improves the robustness of planet detection
\citep[e.g.,][]{haywood_planets_2014,rajpaul_gaussian_2015,faria_uncovering_2016}.
Correlated noise also affects the determination of a planet's parameters
and can induce, in particular, spurious eccentricities
when it is not properly accounted for \citep[e.g.,][]{hara_bias_2019}.

In many cases, the physical processes inducing correlated noise cannot
be modeled precisely but qualitative properties, typical timescales, and amplitudes can be estimated.
Thus, a common approach is to use simple parametric noise models.
For a time series of size $n$ with observations taken at times $t_i$ ($i<n$),
the covariance matrix of the noise is typically modeled as:
\begin{equation}
  \label{eq:defCK}
  C_{i,j} = \delta_{i,j}\sigma_i^2 + K(t_i, t_j),
\end{equation}
where $\sigma_i$ are individual errorbars (e.g., photon noise)
and $K$ is the kernel of the correlated noise.
The noise is often assumed to be stationary, such that $K(t_i,t_j)$ only depends on $|t_i-t_j|$,
\begin{equation}
  K(t_i, t_j) = k(|t_i-t_j|).
\end{equation}
A simple, widespread model assumes the correlation to decrease
exponentially with time, with a timescale of $\tau$,
\begin{equation}
  k(\Delta t) = \sigma_\mathrm{corr.}^2 \expo{-\frac{\Delta t}{\tau}},
\end{equation}
but it is sometimes chosen to decrease as a squared exponential
\citep{schwarzenberg-czerny_accuracy_1991},
\begin{equation}
  k(\Delta t) = \sigma_\mathrm{corr.}^2 \expo{-\frac{\Delta t^2}{2\tau^2}},
\end{equation}
or other similar functions.
Slightly more complex models have also been proposed, for instance,
quasiperiodic kernels, such as
that of \citet{haywood_planets_2014}
\begin{equation}
  \label{eq:haywood}
  k(\Delta t) = \sigma_\mathrm{corr.}^2 \exp\left(
    -\frac{\Delta t^2}{2\tau^2}
    - \frac{2}{\eta}\sin^2\left(\frac{\pi\Delta t}{P_\mathrm{rot.}}\right)
    \right),
\end{equation}
which allow for a more flexible modeling of the underlying physical processes.

In the case of a poorly understood noise source,
the choice of a kernel is somewhat arbitrary
but nonetheless, it should be governed by the qualitative properties
that the noise is expected to present (typical timescales, periodicities, etc.).
For instance, quasiperiodic kernels are well-suited to model the radial velocity signal
induced by stellar spots coming in and out of view due to the rotation of the star
\citep[see][]{haywood_planets_2014}.
Even if the connection to the exact physics of the process is loose,
the qualitative properties of quasiperiodic kernels are sufficient
to bring a significant improvement in detection reliability.

While correlated noise models improve detection robustness,
they might be prohibitive in terms of
computational cost and memory footprint.
Indeed, for a dataset of size $n$,
the covariance matrix of the noise is of size $n \times n$.
In the general case, the memory footprint of storing $C$ is thus $\O\left(n^2\right)$.
Then some operations must be performed with this matrix to compute useful quantities
(such as the $\chi^2$ or the likelihood of a model).
The computational cost of these operations
(e.g., inversion, dot product, determinant)
typically scales as $\O\left(n^2\right)$ to $\O\left(n^3\right)$ in the general case.
These scalings make a correct modeling of the noise intractable for large datasets.
To address this issue,
\citet{ambikasaran_generalized_2015} and \citet{foreman-mackey_fast_2017}
proposed a flexible parametric noise model,
which allow a linear scaling
of the memory footprint and computational cost of the correlated noise.
This so-called \celerite{} model
is capable of handling a mixture of quasiperiodic covariance kernels of the form:
\begin{equation}
  \label{eq:celerite}
  k(\Delta t) = \sum_{s<n_\mathrm{c}}
  \left(a_s \cos(\nu_s\Delta t) + b_s \sin(\nu_s\Delta t) \right) \expo{-\lambda_s\Delta t},
\end{equation}
where $n_c$ is an arbitrarily high number of components in the model.
This model has the property to be semiseparable,
which allows a scaling of the computational cost as $\O\left(n n_c^2\right)$
\citep{ambikasaran_generalized_2015}.
It is similar to the quasiperiodic kernel of \citet{haywood_planets_2014}
which is detailed in Eq.~(\ref{eq:haywood}).
The \celerite{} model is well-suited to represent stellar signals
modulated by the rotation period of the star \citep[e.g.,][]{foreman-mackey_fast_2017}.
It has been used, in particular,
for the analysis of radial velocity and photometric time series.

The star is not the only source of noise in the data.
Instruments also introduce a correlated signature.
For instance, for precise radial velocity time series (and in other fields),
the instrument must be calibrated periodically, typically once per night.
Several scientific measurements might use the same calibration
and, therefore, share the same calibration noise.
The covariance matrix of the calibration noise is then block-diagonal
with the blocks corresponding to each calibration (each night).
This calibration noise is not stationary and, thus, it is not well
represented by the \celerite{} model (see Eq.~(\ref{eq:celerite})).
More generally, when considering various sources of noise together,
the complete covariance matrix might present
quasiperiodic components
and sparse (block diagonal, banded, etc.) components.
While efficient dedicated algorithms exist for both quasiperiodic (semiseparable)
and sparse covariance matrices,
they cannot be applied in a straightforward way for a mixture of both.

In this article, we extend the method described by \citet{foreman-mackey_fast_2017}
to correlated noise with a semiseparable component
plus a sparse component.
We introduce the notion of \leaf{} matrices, a general class of sparse, "close to diagonal"
symmetric matrices
encompassing banded, block-diagonal, staircase matrices, etc.
Our complete model, which we call the \spleaf{} model,
is the sum of a semiseparable component and a \leaf{} component.

In Sect.~\ref{sec:spleaf}, we present the \spleaf{} correlated noise model and dedicated algorithms.
In Sect.~\ref{sec:rv}, we illustrate our methods using the HARPS radial velocities of \object{HD~136352}.
We discuss our results in Sect.~\ref{sec:conclusion}.
We provide an open-source reference implementation of \spleaf{} matrices and related algorithms
as a C library with python wrappers, available at \spleafURL.

\section{The \spleaf{} noise model}
\label{sec:spleaf}

The likelihood (i.e., the probability of the data assuming a given model is correct)
is a common tool for assessing the agreement of a given model to a dataset.
In a Bayesian approach, the quantity of interest is the posterior probability
(probability of a model given the data),
but the computation of the likelihood is still required as an intermediate step.
In this section, we describe the \spleaf{} noise model and dedicated algorithms
which allow, in particular, for the efficient computation of the likelihood and its derivatives.

In Sect.~\ref{sec:generalcase},
we introduce notations and describe the computation of the likelihood in the general case.
We define \spleaf{} matrices in Sect.~\ref{sec:spleafmat},
and we present dedicated algorithms for \spleaf{} matrices
in Sect.~\ref{sec:likelihoodspleaf}.

\subsection{Likelihood computation, general case}
\label{sec:generalcase}

Let us assume that a given dataset $y_i$ ($0\leq i < n$)
can be modeled with a deterministic component (the model) $m$ with parameters $\theta$,
and a correlated Gaussian noise component $\epsilon$ with parameters $\alpha$:
\begin{align}
  y_i & = m_i(\theta) + \epsilon_i,\nonumber\\
  \epsilon & \sim \G(0, C(\alpha)).
\end{align}
The log-likelihood of a given set of parameters ($\theta, \alpha$) is read as:
\begin{align}
  \label{eq:loglike}
  \ln\L(\theta, \alpha) & = \ln p(y|\theta, \alpha)\nonumber                                                   \\
                        & = -\frac{1}{2} \Big(y-m(\theta)\Big)\t C^{-1}(\alpha) \Big(y-m(\theta)\Big)\nonumber \\
                        & \quad -\frac{1}{2} \ln\det \Big(2\pi C(\alpha)\Big),
\end{align}
where $C(\alpha)$ is the $n\times n$ covariance matrix of the correlated noise $\epsilon$.

The computational cost of evaluating the log-likelihood obviously depends
on the cost of evaluating the model $m(\theta)$.
However, once the model is obtained, we still have to compute
the $\chi^2 = r\t C^{-1} r$ (where $r$ represents the residuals, $r = y-m$)
and the determinant of $C$.

An efficient and robust way to compute the log-likelihood in the general case is
to compute the Cholesky decomposition of $C$ as an intermediate step.
By definition, the covariance matrix $C$ is symmetric, positive, and definite.
It could, in principle, be singular (only semi-definite) but this would mean
that some almost-certain affine relation exists in the noise component.
This almost-certain relation could thus be included in the deterministic part of the model.
Assuming $C$ to be invertible (non-singular), its Cholesky decomposition can be read as:
\begin{equation}
  \label{eq:cholesky}
  C = L D L\t,
\end{equation}
where $D$ is diagonal and $L$ is lower triangular with ones on the diagonal.
The classical Cholesky decomposition is actually $C=\Lambda \Lambda\t$,
where $\Lambda = L\sqrt{D}$ is also lower triangular.
However, we use the alternative form of Eq.~(\ref{eq:cholesky}) throughout the article
since this notation is more convenient in our case.
The computational cost of the Cholesky decomposition is $\O\left(n^3\right)$
in the general case.
Once the Cholesky decomposition is obtained,
the determinant of $C$ is easily computed (in $\O\left(n\right)$)
since $\ln\det C = \ln\det D = \displaystyle\sum_i\ln D_i$.
The computation of the $\chi^2$ is performed in $\O\left(n^2\right)$
in the general case by first solving
$u = L^{-1} r$ and then computing $u\t D u$.

\subsection{Symmetric \spleaf{} covariance matrix}
\label{sec:spleafmat}

A common method for improving the computational cost and memory footprint of
correlated noise models is to obtain a sparse representation of the
covariance matrix and to then use dedicated algorithms for
solving, computing the determinant, and other functions that make use of this sparsity.
For instance, dedicated representations and algorithms
for banded matrices, block-diagonal matrices, etc., exist,
allowing for the linear scaling in $n$ of the computational cost and footprint of the model
($\O\left(\alpha n\right)$, where $\alpha$ depends on the bandwidth, block size, etc.).
In \citet{delisle_harps_2018}, the covariance matrix was truncated and approximated by a banded matrix.
This representation improved the computational speed of the Monte Carlo Markov Chain (MCMC) algorithm
used to compute the posterior densities of the orbital elements and noise parameters.

\subsubsection{\leaf{} matrix}

\begin{figure}
  \centering
  \def\leafn{9}
  \def\leafb{0,1,1,2,1,4,1,3,2}
  \def\leafidemo{8}
  \newlength\leafncm
  \setlength\leafncm{\leafn cm}
  \tikzmath{\leafscale = 0.95\linewidth/\leafncm;}
  \begin{tikzpicture}[scale=\leafscale]
    \draw[opacity=0] (0.5,\leafn+0.5) rectangle (\leafn+0.5,0.5);
    \draw[line width=1, dotted] (1,1) grid (\leafn,\leafn);
    \draw[line width=3] (1,\leafn) -- (\leafn,1);
    \foreach \bi [count=\i] in \leafb {
      \draw[line width=3] (\i,\leafn+1-\i) -- (\i-\bi,\leafn+1-\i);
      \draw[line width=3] (\i,\leafn+1-\i) -- (\i,\leafn+1-\i+\bi);
      \foreach \delta in {0,...,\bi} {
          \fill (\i-\delta,\leafn+1-\i) circle (0.15);
          \fill (\i,\leafn+1-\i+\delta) circle (0.15);
        }
      \ifthenelse{\i=\leafidemo} {
        \draw[line width=1, dotted] (1,\leafn+1-\i) -- (0.7,\leafn+1-\i) node[left] {$i$};
        \draw[line width=1, dotted] (\i-\bi,1) -- (\i-\bi,0.7) node[below] {$j=i-b_i$};
        \draw[line width=2, color=red, decorate, decoration={brace,amplitude=5pt}]
        (\i+0.1,\leafn+1-\i-0.15) -- (\i-\bi-0.1,\leafn+1-\i-0.15) node[midway, below, yshift=-6pt, fill=white]{non-zeros};
        \draw[line width=2, color=blue, decorate, decoration={brace,amplitude=5pt}]
        (\i-\bi-1+0.1,\leafn+1-\i-0.15) -- (0.9,\leafn+1-\i-0.15) node[midway, below, yshift=-6pt, fill=white]{zeros};
      }
    }
  \end{tikzpicture}
  \caption{Sketch of a symmetric \leaf{} matrix.}
  \label{fig:leaf}
\end{figure}
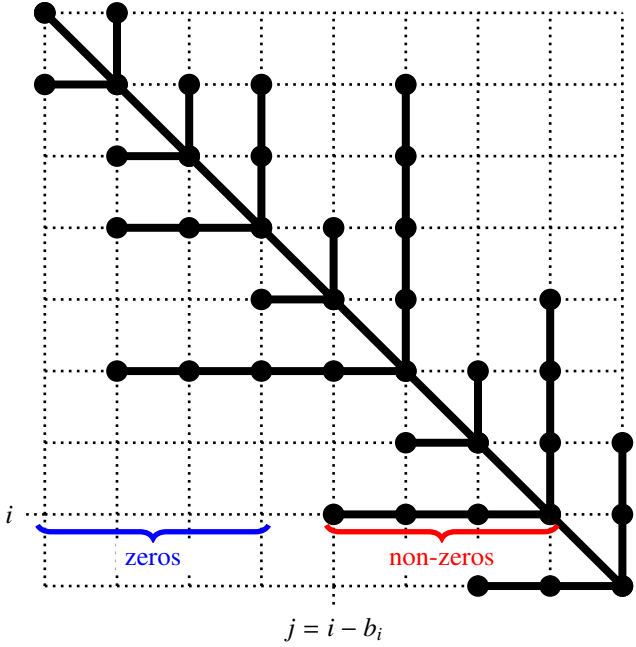

Here we introduce a general class of sparse, "close to diagonal" matrices,
called \leaf{} matrices,
that encompasses banded, block-diagonal, staircase matrices, etc.
A symmetric \leaf{} matrix $F$ must verify:
\begin{equation}
  \label{eq:leaf}
  F_{i,j} = F_{j,i} = 0 \quad \mathrm{for}\ j < i - b_i,
\end{equation}
where $b_i$ is the number of non-zero entries left to the diagonal at line $i$.
A sketch of a symmetric \leaf{} matrix is shown in Fig.~\ref{fig:leaf}.

\subsubsection{Semiseparable matrix}

For efficient computations (typically linear in $n$),
the covariance matrix does not need to be sparse itself,
but it should be expressed as a function of
sparse matrices (sum, product, inverse, etc.).
For instance, \citet{rybicki_class_1995} showed that exponential matrices,
defined as:
\begin{equation}
  \label{eq:expomat}
  C_{i,j} = \expo{-\lambda \Delta t_{i,j}},
\end{equation}
with $\Delta t_{i,j} = |t_i-t_j|$,
possess a tridiagonal inverse $T$ which can be computed directly,
without requiring to compute $C$ first \citep[see][]{rybicki_class_1995}.
While the covariance matrix $C$ is not sparse,
using the property $C = T^{-1}$ allows for a very efficient (i.e., in $\O\left(n\right)$)
representation and computation.
These exponential matrices (as defined in Eq.~(\ref{eq:expomat}))
are also a particular example of semiseparable matrices
which makes it possible to obtain another sparse representation.
Indeed, assuming $t$ to be ordered increasingly,
and defining $u = \expo{-\lambda t}$ and $v = \expo{\lambda t}$ (vectors of size $n$),
$C$ can be decomposed as:
\begin{equation}
  C = \id + \tril\left(u v\t\right) + \triu\left(v u\t\right),
\end{equation}
where $\tril$ (respectively $\triu$) stands for
the strictly lower (respectively upper) triangular part.
The two sparse representations of the exponential matrix of Eq.~(\ref{eq:expomat})
(i.e., tridiagonal inverse and semiseparable form)
are actually linked one to the other,
since the inverse of invertible tridiagonal matrices are rank one semiseparable matrices
and vice-versa \citep[e.g.,][and references therein]{vandebril_bibliography_2005}.

More generally, a symmetric semiseparable matrix is defined as:
\begin{equation}
  \label{eq:spleafmat}
  C = \diag(A) + \tril\left(U V\t\right) + \triu\left(V U\t\right),
\end{equation}
where $\diag(A)$ is the diagonal matrix built from the vector $A$ (size $n$),
$U$, and $V$ are $(n\times r)$ matrices,
and $r$ is the rank of the semiseparable matrix $C$.
Semiseparable matrices can represent a large class of correlated noise models.
For instance, the \celerite{} model (see Eq.~(\ref{eq:celerite})) proposed by \citet{foreman-mackey_fast_2017}
can be represented as a semiseparable matrix of rank $r = 2 n_\mathrm{c}$, with:
\begin{align}
  \label{eq:celeritedecomp}
  A_i                  & = \sigma_i^2 + \sum_{s<n_\mathrm{c}}a_s,\nonumber                                        \\
  U_{i,s}              & = \expo{-\lambda_s t_i} \left(a_s \cos(\nu_s t_i) + b_s \sin(\nu_s t_i)\right),\nonumber \\
  U_{i,n_\mathrm{c}+s} & = \expo{-\lambda_s t_i} \left(a_s \sin(\nu_s t_i) - b_s\cos(\nu_s t_i)\right),\nonumber  \\
  V_{i,s}              & = \expo{\lambda_s t_i} \cos(\nu_s t_i),\nonumber                                         \\
  V_{i,n_\mathrm{c}+s} & = \expo{\lambda_s t_i} \sin(\nu_s t_i).
\end{align}
The computational cost and memory footprint of a semiseparable noise model are linear in $n$
\citep[footprint in $\O\left(rn\right)$ and cost in $\O\left(r^2n\right)$, see][]{ambikasaran_generalized_2015,foreman-mackey_fast_2017}.

\subsubsection{\spleaf{} matrix}

We define a \spleaf{} matrix simply as the sum of a semiseparable and a \leaf{} matrix.
A symmetric \spleaf{} matrix takes, thus, the form of:
\begin{equation}
  \label{eq:spleaf}
  C = \diag(A) + \tril\left(U V\t\right) + \triu\left(V U\t\right) + F,
\end{equation}
where
$A$ is a vector of size $n$ representing the diagonal part of $C$,
$U$, and $V$ are $n\times r$ matrices representing the symmetric semiseparable part of $C$,
and $F$ is the symmetric \leaf{} part of $C$, as defined in Eq.~(\ref{eq:leaf}).
Since the diagonal part of $C$ is represented by the vector $A$,
we assume the diagonal of $F$ to be filled with zeros.
As in Eq.~(\ref{eq:leaf}),
we denote by $b_i$ the number of non-zero entries left to the diagonal,
at line $i$ of $F$ (see also Fig.~\ref{fig:leaf}).
The sparse matrix $F$ can thus be stored in a compact way
(i.e., storing only non-zero entries, and using its symmetry)
with $\mean{b}n$ values.
The memory footprint of the \spleaf{} model scales as $\O\left(\left(r+\mean{b}\right)n\right)$
and the computational cost as $\O\left(\left(r^2+r\mean{b}+\mean{b^2}\right)n\right)$,
where $r$ is the number of components in the semiseparable part
and for any vector $x$, $\mean{x}$ stands for the mean of $x$.

\subsection{Likelihood computation with \spleaf{} matrices}
\label{sec:likelihoodspleaf}

\subsubsection{Cholesky decomposition}
\label{sec:choleskyspleaf}

We then look for a sparse representation and an efficient computation
of the matrices $D$ and $L$ involved in the
Cholesky decomposition (see Eq.~(\ref{eq:cholesky})) of $C$ as defined by Eq.~(\ref{eq:spleaf}).
In the case $F=0$, \citet{foreman-mackey_fast_2017} showed that $L$ can be written as:
\begin{equation}
  \label{eq:Lsep}
  L = \id + \tril\left(U W\t\right),
\end{equation}
where $W$ is a new $n\times r$ matrix which need to be determined.
In the case $F\neq 0$, this decomposition does not hold but we can prove that there exist
a $n\times r$ matrix $W$ and
a strictly lower triangular \leaf{} matrix $G$ with the same shape as $F$ (i.e., same values of $b_i$),
such that:
\begin{equation}
  \label{eq:L}
  L = \id + \tril\left(U W\t\right) + G.
\end{equation}

Let us first simply assume that $G$ is strictly lower triangular (not necessarily \leaf{}).
In this case, the decomposition is degenerated but always exists.
Replacing $L$ by the expression of Eq.~(\ref{eq:L}) in the Cholesky decomposition
of $C$ (Eq.~(\ref{eq:cholesky})) and equating it to Eq.~(\ref{eq:spleaf}),
we obtain (for $j < i$):
\begin{align}
  \label{eq:Cii}
  C_{i,i} & = A_i
  = D_i + \sum_{k<i} \left(\sum_{s} U_{i,s} W_{k,s} + G_{i,k}\right)^2 D_k\nonumber                     \\
          & = D_i + \sum_{s} U_{i,s} \left(\sum_{t} S_{i,s,t} U_{i,t} + 2 Z_{i,i,s}\right)
  + \sum_{k<i} G_{i,k}^2 D_k,                                                                           \\
  \label{eq:Cij}
  C_{i,j} & = \sum_{s} U_{i,s} V_{j,s} + F_{i,j}\nonumber                                               \\
          & = \left(\sum_{s} U_{i,s} W_{j,s} + G_{i,j}\right) D_j\nonumber                              \\
          & \quad + \sum_{k<j} \left(\sum_{s} U_{i,s} W_{k,s} + G_{i,k}\right)
  \left(\sum_{t} U_{j,t} W_{k,t} + G_{j,k}\right) D_k\nonumber                                          \\
          & = \sum_{s} U_{i,s}\left(W_{j,s}D_j + \sum_{t} U_{j,t} S_{j,s,t} + Z_{j,j,s}\right)\nonumber \\
          & \quad + G_{i,j} D_j + \sum_{k<j} G_{i,k} G_{j,k} D_k
  + \sum_{s}U_{j,s} Z_{i,j,s},
\end{align}
where $S$ is defined following \citet{foreman-mackey_fast_2017},
\begin{equation}
  \label{eq:defS}
  S_{i,s,t} = \sum_{k<i} W_{k,s} D_k W_{k,t},
\end{equation}
and $Z$ is defined as:
\begin{equation}
  \label{eq:defZ}
  Z_{i,j,s} = \sum_{k<j} G_{i,k} D_k W_{k,s}.
\end{equation}
We then break the degeneracy in the expression of $L$
by identifying the terms in front of $U_{i,s}$ in Eq.~(\ref{eq:Cij}).
Thus we obtain:
\begin{align}
  V_{j,s} & = W_{j,s}D_j + \sum_{t} U_{j,t} S_{j,s,t} + Z_{j,j,s},\nonumber \\
  F_{i,j} & = G_{i,j} D_j + \sum_{k<j} G_{i,k} G_{j,k} D_k
  + \sum_{s}U_{j,s} Z_{i,j,s}.
\end{align}
We deduce the following expressions for $D$, $W$, and $G$ (for $j<i$):
\begin{align}
  \label{eq:recDfull}
  D_i     & = A_i - \sum_{s} U_{i,s} \left(\sum_{t} S_{i,s,t} U_{i,t} + 2 Z_{i,i,s}\right)
  - \sum_{k<i} G_{i,k}^2 D_k,                                                              \\
  \label{eq:recWfull}
  W_{i,s} & = \frac{1}{D_i}\left(V_{i,s} - \sum_{t} S_{i,s,t} U_{i,t} - Z_{i,i,s}\right),  \\
  \label{eq:recGfull}
  G_{i,j} & = \frac{1}{D_j}\left(F_{i,j} - \sum_{k<j} G_{i,k} G_{j,k} D_k
  - \sum_{s}U_{j,s} Z_{i,j,s}\right).
\end{align}
From Eqs.~(\ref{eq:defZ}) and (\ref{eq:recGfull}),
we can check by induction that $G_{i,j} = Z_{i,j,s} = 0$ for $j < i-b_i$.
Therefore, $G$ and $Z$ have the same \leaf{} shape as $F$,
which proves that the decomposition of Eq.~(\ref{eq:L}) always exists.

Using this property, we are able to compute compact recursion formulas
for the expression of $S$, $Z$, $G$, $D$, and $W$.
We find that for increasing values of $i$
and increasing values of $j$ at $i$ fixed
(with $i-b_i\leq j<i$):
\begin{align}
  \label{eq:recS}
  & S_{0,s,t} = 0,\nonumber                                                                                     \\
  & S_{i,s,t} = S_{i-1,s,t} + W_{i-1,s} D_{i-1} W_{i-1,t}\quad (i>0),                                           \\
  \label{eq:recZ}
  & Z_{i,i-b_i,s} = 0,\nonumber                                                                                 \\
  & Z_{i,j,s} = Z_{i,j-1,s} + G_{i,j-1} D_{j-1} W_{j-1,s}\quad (j>i-b_i),                                       \\
  \label{eq:recG}
  & G_{i,j} = \frac{1}{D_j}\left(F_{i,j} - \sum_{k=\max(i-b_i, j-b_j)}^{j-1}\hspace{-6.5mm} G_{i,k} G_{j,k} D_k
  - \sum_{s}U_{j,s} Z_{i,j,s}\right),                                                                            \\
  \label{eq:recD}
  & D_i = A_i - \sum_{s} U_{i,s} \left(\sum_{t} S_{i,s,t} U_{i,t} + 2 Z_{i,i,s}\right)
  - \sum_{k=i-b_i}^{i-1}\hspace{-1mm} G_{i,k}^2 D_k,                                                             \\
  \label{eq:recW}
  & W_{i,s} = \frac{1}{D_i}\left(V_{i,s} - \sum_{t} S_{i,s,t} U_{i,t} - Z_{i,i,s}\right).
\end{align}
While $S$ is a $n\times r\times r$ tensor, it is not necessary to keep all its values in memory,
and $S$ can be stored as $r\times r$ matrix which is updated in place for increasing values of $i$.
The same reasoning holds for $Z$, which can be stored as a vector of size $r$,
and updated for increasing values of $i$ and $j$.
However, if the backpropagation of the gradient is required,
all the values of $S$ and $Z$ should be stored for reasons of stability and performance
(see Sect.~\ref{sec:backprop}).
In this case, the memory footprint of the \spleaf{} model increases but remains linear in $n$
(i.e., $\O\left(\left(r+\mean{b}\right)rn\right)$ instead of $\O\left(\left(r+\mean{b}\right)n\right)$).

\subsubsection{Computing the determinant and solving}
\label{sec:detsolve}

As explained in Sect.~\ref{sec:generalcase}, once the Cholesky decomposition
of the covariance matrix $C$ is known, we need to compute its determinant
and solve for $x = L^{-1} y$ to compute the likelihood of a set of parameters.
The determinant is trivially obtained in $\O(n)$ operations,
\begin{equation}
  \ln\det(C) = \ln\det(D) = \sum_i \ln D_i.
\end{equation}
We can then describe how to solve for $x = L^{-1} y$ (with $L$ defined as in Eq.~(\ref{eq:L})).
Since $y = Lx$, we have:
\begin{align}
  y_i & = x_i + \sum_{j<i} \left(\sum_s U_{i,s}  W_{j,s} + G_{i,j}\right) x_j\nonumber \\
      & = x_i + \sum_s U_{i,s} f_{i,s} + \sum_{j=i-b_i}^{i-1} G_{i,j} x_j,
\end{align}
with $f$ defined as in \citet{foreman-mackey_fast_2017},
\begin{equation}
  f_{i,s} = \sum_{j<i} W_{j,s} x_j.
\end{equation}
We thus obtain the following recursion formulas for increasing values of $i$:
\begin{align}
  \label{eq:solveLf}
  f_{0,s} & = 0,\nonumber                                                      \\
  f_{i,s} & = f_{i-1,s} + W_{i-1,s} x_{i-1}\quad (i>0),                        \\
  \label{eq:solveLx}
  x_i     & = y_i - \sum_s U_{i,s} f_{i,s} - \sum_{j=i-b_i}^{i-1} G_{i,j} x_j.
\end{align}
As for the Cholesky factorization, the values of $f$ can be stored in a vector of size $r$
and updated in place for increasing values of $i$, except in the case where the backpropagation
of the gradient is required (see Sect.~\ref{sec:backprop}).
The computational cost of this solving is in $\O\left(\left(r+\mean{b}\right)n\right)$.

While it is not needed in the calculation of the likelihood,
the computation of the dot product $y = Lx$
is very similar to the solving problem ($x = L^{-1}y$).
For increasing values of $i$, we compute:
\begin{align}
  \label{eq:dotLf}
  f_{0,s} & = 0,\nonumber                                                      \\
  f_{i,s} & = f_{i-1,s} + W_{i-1,s} x_{i-1}\quad (i>0),                        \\
  \label{eq:dotLy}
  y_i     & = x_i + \sum_s U_{i,s} f_{i,s} + \sum_{j=i-b_i}^{i-1} G_{i,j} x_j.
\end{align}
Similar recursion formulas for the dot product $y = L\t x$
and the solving of $x = L\mt y$ are easily obtained.

\subsubsection{Overflows and preconditioning}
\label{sec:overflow}

As noted by \citet{ambikasaran_generalized_2015,foreman-mackey_fast_2017},
a naive computer implementation of exponential semiseparable matrices can lead to
numerical underflows and overflows.
Indeed, the separation of the exponential $\expo{-\lambda_s |t_i-t_j|}$
in $U_{i,s} = \expo{-\lambda_s t_i}$ and $V_{j,s} = \expo{\lambda_s t_j}$
exhibits very interesting theoretical properties (semiseparable matrix)
but in practical applications, $\lambda_s t_i$ and $\lambda_s t_j$
can reach values that are much larger than $\lambda_s |t_i-t_j|$,
which causes underflows for $U$ and overflows for $V$.

To circumvent this numerical issue, we follow \citet{foreman-mackey_fast_2017}
and introduce the $(n-1)\times r$ preconditioning matrix $\phi$, and the preconditioned
matrices $\tilde{U}$ and $\tilde{V}$, such that
\begin{equation}
  U_{i,s} V_{j,s} = \tilde{U}_{i,s} \tilde{V}_{j,s} \prod_{k=j}^{i-1} \phi_{k,s}.
\end{equation}
For instance, in the case of the \celerite{} model -- Eqs.~(\ref{eq:celerite}) and (\ref{eq:celeritedecomp}) --
\citet{foreman-mackey_fast_2017} proposed the following preconditioning:
\begin{align}
  \tilde{U}_{i,s}              & = a_s \cos(\nu_s t_i) + b_s \sin(\nu_s t_i),\nonumber       \\
  \tilde{U}_{i,n_\mathrm{c}+s} & = a_s \sin(\nu_s t_i) - b_s\cos(\nu_s t_i),\nonumber        \\
  \tilde{V}_{i,s}              & = \cos(\nu_s t_i),\nonumber                                 \\
  \tilde{V}_{i,n_\mathrm{c}+s} & = \sin(\nu_s t_i),\nonumber                                 \\
  \phi_{i,s}                   & = \phi_{i,n_\mathrm{c}+s} = \expo{-\lambda_s(t_{i+1}-t_i)},
\end{align}
which avoids the computation of exponentials with large exponents.
All the algorithms presented above (Cholesky decomposition, dot product and solving)
can be adapted to take into account this preconditioning.
We refer the reader to Appendix~\ref{sec:overflowdetails} for more details.

\subsubsection{Efficient computation of the likelihood derivatives}
\label{sec:backprop}

Once the model is chosen,
we typically need to determine a point estimate or the posterior distribution of the parameters.
In order to use efficient optimization or exploration algorithms,
it might be useful to compute the gradient of the log-likelihood (Eq.~(\ref{eq:loglike}))
with respect to the model parameters ($\theta$) and the noise parameters ($\alpha$).
\citet{foreman-mackey_scalable_2018} provided gradient backpropagation algorithms
for the Cholesky decomposition, dot product, and solving problem
in the case of semiseparable matrices ($F=0$ in our notations).
These algorithms allow to very efficiently compute
\citep[in $\O\left(r^2n\right)$, see][]{foreman-mackey_scalable_2018}
the gradient of the log-likelihood using analytical formulas.
The generalization of this method to \spleaf{} matrices is straightforward
and we provide more details in Appendix~\ref{sec:backpropdetails}.

\section{Application to the analysis of radial velocities}
\label{sec:rv}

In this section, we illustrate the use of the \spleaf{} noise model
by reanalyzing the HARPS radial velocity time series of \object{HD~136352}
\citep[see][]{udry_harps_2019}.
The star \object{HD~136352} is a quiet G4V star known to host three super-Earth planets,
at periods of 11.5824~d, 27.5821~d, and 107.6~d,
and with minimum masses of 4.8, 10.8, and 8.6~$M_\earth$ respectively
\citep[see][]{udry_harps_2019}.
These results were obtained by binning the data and only searching for planets with periods above 1~d.
This is a common practice that allows to damp many instrumental and stellar short-term variations
\citep{dumusque_planetary_2011}.
However, it does not allow us to characterize these short-term variations and to fully correct for them.
Moreover, binning the data could significantly damp the amplitude of short period planets.
Here we reanalyze the raw radial velocities and do not restrict our study to periods above 1~d.

The radial velocities of \object{HD~136352} taken with HARPS consist of 648 points,
taken over almost 11 years (2004-2015),
and spread over 238 distinct nights.
The number of points per night varies between one and ten, with an average of 2.7 points per night.

We describe the different noise models we use for our study in Sect.~\ref{sec:noisemodels}
and present our reanalysis of the \object{HD~136352} system in Sect.~\ref{sec:rvsystem}.

\subsection{Noise models}
\label{sec:noisemodels}

\begin{figure*}
  \centering
  \includegraphics[width=0.95\linewidth]{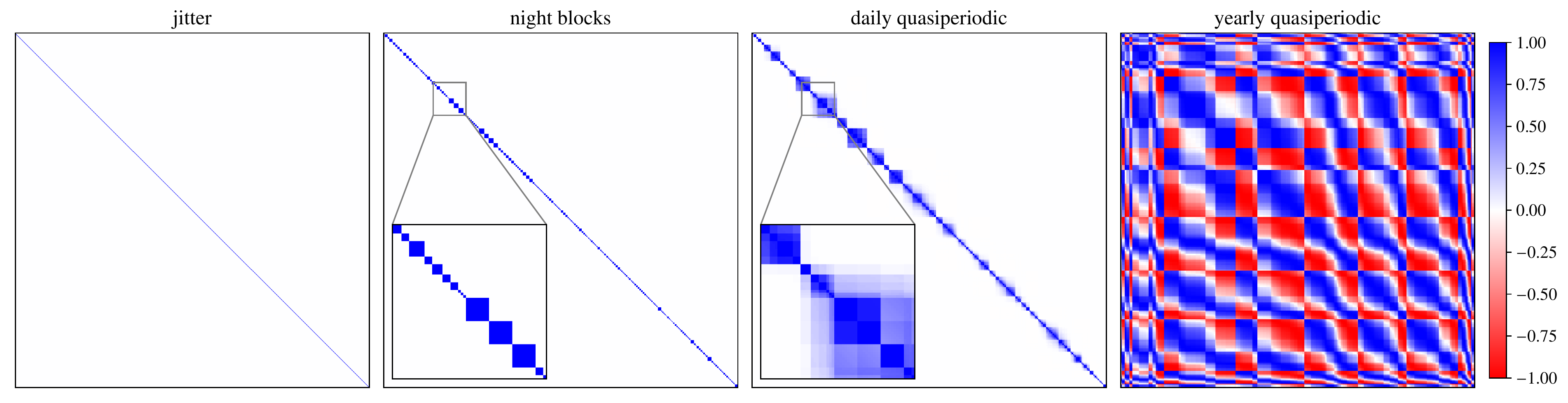}
  \caption{Shapes of the four components of the noise models used for the analysis of the \object{HD~136352} system (Sect.~\ref{sec:rv}).}
  \label{fig:covs_HD136352}
\end{figure*}

To illustrate the role of each component in our \spleaf{} noise model,
we analyze the data using five different noise models:
\begin{enumerate}
  \item diag.: a diagonal matrix, with the observational errorbars $\sigma_i$ plus a jitter term
    ($\sigma_\mathrm{jit.}$) added in quadrature (same value for all data points)
    \begin{equation}
      C_{i,j} = (\sigma_i^2 + \sigma_\mathrm{jit.}^2)\delta_{i,j};
    \end{equation}
  \item bin.: same as diag. but using nightly binned radial velocity data;
  \item \celerite: same as diag. plus quasiperiodic terms at 1~d and 1~yr,
  \begin{align}
    C_{i,j} & = (\sigma_i^2 + \sigma_\mathrm{jit.}^2)\delta_{i,j} \nonumber\\
    &+ \sigma_\mathrm{d}^2 \expo{-0.1|t_i-t_j|} \cos\left(2\pi\frac{t_i-t_j}{1~\mathrm{d}}\right)\nonumber\\
    &+ \sigma_\mathrm{yr}^2 \cos\left(2\pi\frac{t_i-t_j}{1~\mathrm{yr}}\right);
  \end{align}
    \item \leaf: same as diag. but the estimated calibration error $\sigma_{b_i}$
    (which is part of the observational error $\sigma_i$)
    is shared by night blocks (identified by $b_i$),
    and an additional calibration error term ($\sigma_\mathrm{cal.}$) is added in quadrature to these blocks
    (same value for all blocks),
    \begin{align}
      C_{i,j} & = (\sigma_i^2-\sigma_{b_i}^2 + \sigma_\mathrm{jit.}^2)\delta_{i,j}\nonumber\\
      &+ (\sigma_{b_i}^2 + \sigma_\mathrm{cal.}^2)\delta_{b_i,b_j};
    \end{align}
  \item \spleaf: same as \leaf{} plus the two quasiperiodic terms at 1~d and 1~yr as in the \celerite{} model,
  \begin{align}
    C_{i,j} & = (\sigma_i^2-\sigma_{b_i}^2 + \sigma_\mathrm{jit.}^2)\delta_{i,j}\nonumber\\
    &+ (\sigma_{b_i}^2 + \sigma_\mathrm{cal.}^2)\delta_{b_i,b_j}\nonumber\\
    &+ \sigma_\mathrm{d}^2 \expo{-0.1|t_i-t_j|} \cos\left(2\pi\frac{t_i-t_j}{1~\mathrm{d}}\right)\nonumber\\
    &+ \sigma_\mathrm{yr}^2 \cos\left(2\pi\frac{t_i-t_j}{1~\mathrm{yr}}\right).
  \end{align}
\end{enumerate}
The quasiperiodic terms of the \celerite{} and \spleaf{} models are modeled
according to Eq.~(\ref{eq:celerite}) and could represent instrumental systematics
\citep[CCD stitching, wavelength solution instabilities,
incorrect BERV correction, incorrect airmass corrections, etc.; see][]{dumusque_characterization_2015}.
The HARPS radial velocities of \object{HD~136352} are already corrected from the CCD stitching issue
using the method of \citet{dumusque_characterization_2015},
but remaining systematics could still be present.
The amplitudes of the cosines ($a_s$ in Eq.~(\ref{eq:celerite}))
are noted $\sigma_\mathrm{d}$ and $\sigma_\mathrm{yr}$.
For the sake of simplicity,
we fix the amplitudes of the sines to zero ($b_s=0$),
such that the correlation is always maximum for $\Delta t = 0$ (see Eq.~(\ref{eq:celerite})).
The exponential decay timescale is fixed to $10~d$ for the daily term ($\lambda_\mathrm{d} = 1/10$)
and is infinite for the yearly term ($\lambda_\mathrm{yr} = 0$).

The noise parameters that remain to be determined are, thus,
$\alpha=(\sigma_\mathrm{jit.}^2, \sigma_\mathrm{cal.}^2, \sigma_\mathrm{d}^2, \sigma_\mathrm{yr}^2)$,
or a subset of it depending on the chosen noise model.
The components of the covariance matrices corresponding to each of these four parameters
are illustrated in Fig.~\ref{fig:covs_HD136352}.
For these illustrations, the matrices are expanded as full $n\times n$ matrices,
but we use their sparse representation (as described in Sect.~\ref{sec:spleaf})
in the following computations.

\subsection{Reanalysis of the HD~136352 system}
\label{sec:rvsystem}

\begin{figure*}
  \centering
  \includegraphics[width=0.9\linewidth]{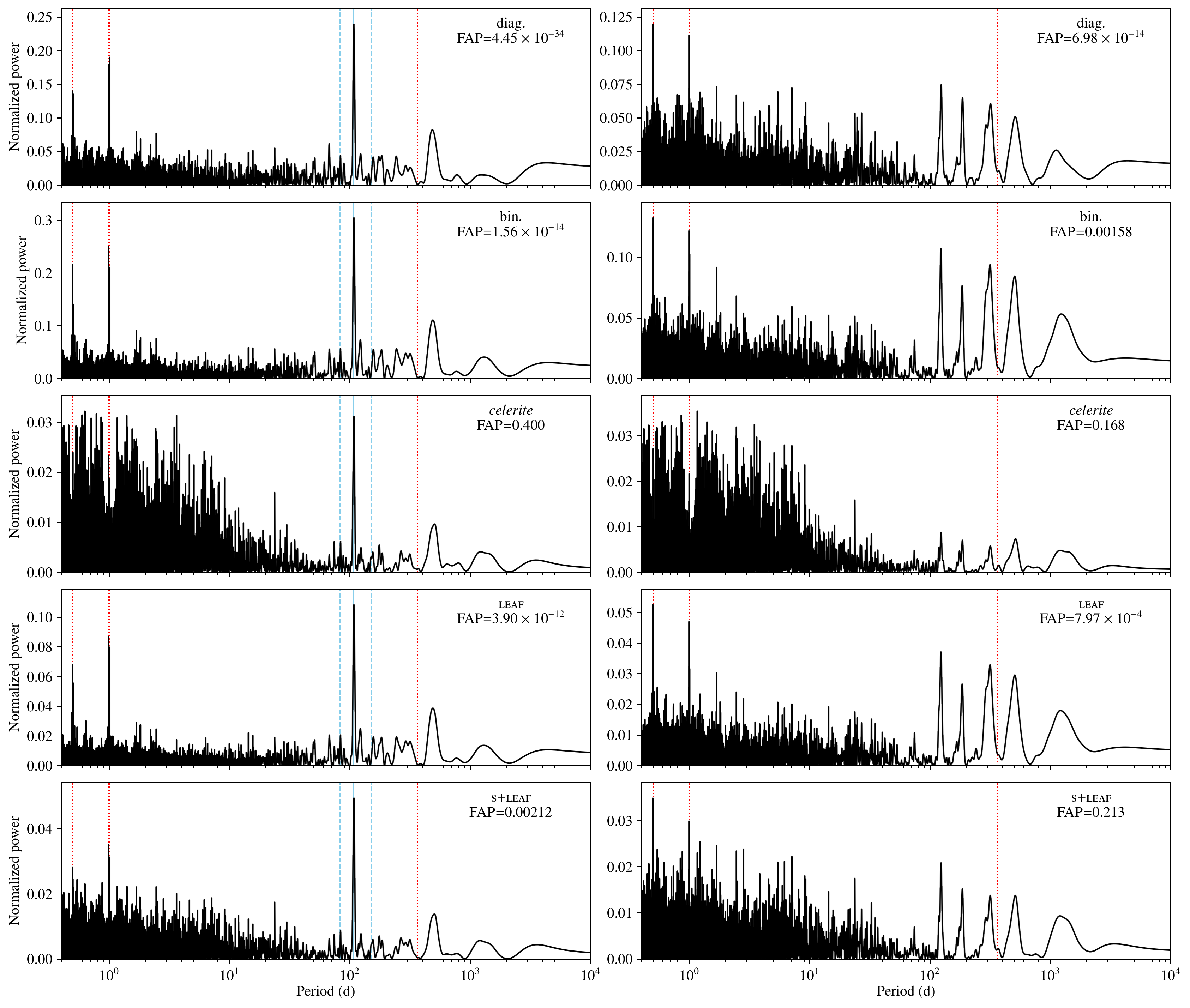}
  \caption{Periodograms of the radial velocity residuals of \object{HD~136352}
    after subtracting the two first planets (at 11.5824~d and 27.5821~d, \textit{left}),
    and after subtracting the three known planets (11.5824~d, 27.5821~d, and 107.6~d \textit{right}),
    for the five noise models defined in Sect.~\ref{sec:noisemodels}.
    The noise parameters are set to the values provided in Table~\ref{tab:noiseparams_HD136352}.
    The vertical blue line highlights the period of the third planet (107.6~d),
    and the dashed blue lines highlight its aliases at 1~yr.
    The dotted vertical red lines highlight 0.5~sd, 1~sd, and 1~yr.
    For the sake of readability, we do not show here the two first periodograms (raw time series
    and after subtracting the first planet),
    since the two first planets (11.5824~d and 27.5821~d)
    are unambiguously detected
    (highest peaks and $\mathrm{FAP} < 10^{-10}$)
    independently of the noise model.
    Assuming that the 107.6~d signal is due to a planet
    while the signals at 0.5~sd, 1~sd, and around 1~yr are due to correlated noise,
    we expect the correct noise model to show a low FAP in the left column
    and a high FAP in the right column.
  }
  \label{fig:perio_HD136352}
\end{figure*}

We analyze the HARPS radial velocity time series of \object{HD~136352}
using each of the five noise models of Sect.~\ref{sec:noisemodels}.
The deterministic part of the model is read as:
\begin{equation}
  \label{eq:deterministicmodel}
  y_i = \gamma + \sum_{p<n_p} K_p \left( \cos(v_p(t) + \omega_p) + e_p\cos\omega_p \right),
\end{equation}
where $\gamma$ is the velocity offset,
$n_p$ is the number of planets,
and, for each planet $p$,
$K_p$ is its semi-amplitude,
$v_p$ its true anomaly,
$e_p$ its eccentricity,
and $\omega_p$ its argument of periastron.
We start our study by considering a model without any planet
and add them gradually, one after the other, by computing a periodogram of the residuals.
At each step of this process, we adjust all the free parameters
(deterministic and noise parameters).
The deterministic parameters (vector $\theta$) are
the offset $\gamma$ and the orbital parmeters
$P$, $K$, $M_0$ (mean anomaly at a reference epoch), $e$, and $\omega$
for each planet included in the model.
The noise parameters $\alpha$ are a subset of
$(\sigma_\mathrm{jit.}^2, \sigma_\mathrm{cal.}^2, \sigma_\mathrm{d}^2, \sigma_\mathrm{yr}^2)$
depending on the chosen noise model.
We use the L-BFGS-B algorithm
\citep{byrd_limited_1995}
to maximize the likelihood (Eq.~(\ref{eq:loglike})) and we make use of
the backpropagation algorithms described in Sect.~\ref{sec:backprop} (see also Appendix~\ref{sec:backpropdetails})
to compute the derivatives of the log-likelihood with respect to the free parameters.
We also use classical analytical expressions for the derivatives of the
Keplerian model (Eq.~(\ref{eq:deterministicmodel}))
with respect to the orbital parameters of the planets.
Then we compute a periodogram of the residuals of this maximum likelihood solution.
The offset $\gamma$ is readjusted for each frequency explored in the periodogram,
but the previous planets and noise parameters are fixed (at the values obtained with the last fit).

We compute the periodograms and associated false alarm probability (FAP) using the analytical method of
\citet{delisle_efficient_2020}, based on the previous work by \citet{baluev_assessing_2008}.
For a frequency $\nu$, we define the normalized power as:
\begin{equation}
  \label{eq:normpower}
  \mathrm{Normalized\ Power\ }(\nu) = \frac{\chi^2_\H - \chi^2_\K(\nu)}{\chi^2_\H},
\end{equation}
which corresponds to the definition of the Generalized Lomb-Scargle periodogram
\citep[GLS, see][]{ferraz-mello_estimation_1981,zechmeister_generalised_2009},
and to $(2/n_\H) z_1(\nu)$ in the notations of \citet{baluev_assessing_2008}
and \citet{delisle_efficient_2020}.
In this definition, $\H$ stands for the base model (only the offset $\gamma$ is adjusted)
and $\K$ stands for the model with frequency $\nu$
($\gamma$ plus the amplitudes of the sine and cosine at frequency $\nu$ are adjusted).
The $\chi^2$ of a model $m(\theta)$ is defined as:
\begin{equation}
  \chi^2 = r^T C^{-1} r,
\end{equation}
where $r$ is the vector of the model residuals ($r = y - m(\theta)$).

The resulting periodograms are shown in Fig.~\ref{fig:perio_HD136352}.
For the sake of readability, we do not show the first two periodograms
since the first two planets (at 11.5824~d and 27.5821~d) are unambiguously detected
(highest peaks and $\mathrm{FAP} < 10^{-10}$)
independently of the noise model.
We additionally provide in Table~\ref{tab:noiseparams_HD136352}
the values of the noise parameters used to compute each of
the periodograms of Fig.~\ref{fig:perio_HD136352}.

\begin{table}
  \caption{Noise parameters adjusted for \object{HD~136352}
    and used to compute the periodograms of Fig.~\ref{fig:perio_HD136352}.}
  \begin{center}
    \footnotesize
    \setlength{\tabcolsep}{1ex}
    \begin{tabular}{c|ccccc}
      \hline
      \hline
      & diag. & bin. & \textit{celerite} & \textsc{leaf} & \textsc{s+leaf} \\
      \hline
      $\sigma_\mathrm{jit.}^2$ & 2.73, 1.94 & 2.46, 1.63 & 0.37, 0.39 & 0.39, 0.39 & 0.39, 0.39 \\
      $\sigma_\mathrm{d}^2$ & -- & -- & 3.23, 2.67 & -- & 1.69, 0.70 \\
      $\sigma_\mathrm{yr}^2$ & -- & -- & 0.00, 0.00 & -- & 0.00, 0.00 \\
      $\sigma_\mathrm{cal.}^2$ & -- & -- & -- & 2.28, 1.45 & 0.56, 0.78 \\
      \hline
    \end{tabular}
  \end{center}
  \tablefoot{For each parameter, the first value corresponds to the left column
    and the second value to the right column of Fig.~\ref{fig:perio_HD136352}.}
  \label{tab:noiseparams_HD136352}
\end{table}

We observe in Fig.~\ref{fig:perio_HD136352} (left column)
that the last planet (HD~136352~d) is well revovered (highest peak and low FAP)
by all models except the \celerite{} model.
With the \celerite{} model, the peak corresponding to the planet is not the highest peak
and the FAP is high (0.4).
We see in Table~\ref{tab:noiseparams_HD136352},
that the amplitude of the daily quasiperiodic term of the \celerite{} model is adjusted to a high value
(3.23~$\mathrm{m}^2/\mathrm{s}^2$).
On the contrary, for the \spleaf{} model, the amplitude of the noise is shared between
the daily quasiperiodic term (1.69~$\mathrm{m}^2/\mathrm{s}^2$) and the calibration noise
(0.56~$\mathrm{m}^2/\mathrm{s}^2$).
It thus seems that the daily quasiperiodic term of the \celerite{} model
is overestimated due to the presence of the unmodeled calibration noise.
This shows that the way the calibration noise is accounted for in the \spleaf{} model
is well suited and does correspond to the behavior of the HARPS instrument.

The periodograms of the residuals of \object{HD~136352}
after subtracting all known planets (Fig.~\ref{fig:perio_HD136352}, right)
do not show any significant peak for the \celerite{} ($\mathrm{FAP} = 0.168$)
and \spleaf{} ($\mathrm{FAP} = 0.213$) models.
On the contrary, the diag., bin., and \leaf{} models
show significant peaks (with a low FAP) around 0.5~sd and 1~sd, as well as around 1~yr
(see Fig.~\ref{fig:perio_HD136352}, right).
These signals could be of planetary origin but are more probably due to instrumental systematics
\citep[CCD stitching, wavelength solution instabilities,
incorrect BERV correction, incorrect airmass corrections, etc.; see][]{dumusque_characterization_2015}.
They could also originate from a combination of stellar correlated noise and aliasing.
These potential systematics are taken into account in the \celerite{} and \spleaf{} models
with the daily and yearly quasiperiodic terms.
While the amplitude of the daily quasiperiodic term is adjusted to significant values
in the \celerite{} and \spleaf{} models,
the amplitude of the yearly quasiperiodic term is completely negligible in both cases
(see Table~\ref{tab:noiseparams_HD136352}).
We performed a similar analysis on the HARPS radial velocities of \object{HD~136352}
without the stitching correction and obtained higher values for the yearly term
($\sigma_\mathrm{yr}^2\approx 0.25\ \mathrm{m}^2/\mathrm{s}^2$).
This highlights the improvements in the radial velocities obtained with this correction.
In the \spleaf{} model, the final levels (after substracting all known planets)
of the daily quasiperiodic term and the calibration term are of the same order of magnitude
(respectively, 0.7 and 0.78~$\mathrm{m}^2/\mathrm{s}^2$, see Table~\ref{tab:noiseparams_HD136352}).
This provides a good illustration of the importance of taking into account both components in the noise model.

The modeling of the systematics using daily and yearly quasiperiodic terms is a rough approximation,
and a further investigation is necessary
to confirm that these signals are instrumental systematics,
to better characterize the systematics for several systems,
to understand the mechanisms that might introduce them,
and to correct for them, ideally directly in the HARPS data reduction software (DRS).
However, this is beyond the scope of this study,
and we simply highlight the ability of the \spleaf{} model to roughly account for these systematics.

\section{Conclusion}
\label{sec:conclusion}

In this article, we present the \spleaf{} correlated noise model.
While in the general case, accounting for correlated noise in a dataset of size $n$ has
a cost of $\O\left(n^3\right)$ and a footprint of $\O\left(n^2\right)$,
the \spleaf{} noise model scales linearly (i.e., in $\O\left(n\right)$).
This linear scaling is made possible by the sparse properties of the \spleaf{} covariance matrices
(see Sect.~\ref{sec:spleaf}).
The \spleaf{} model incorporate a mixture of quasiperiodic components (see Eq.~(\ref{eq:celerite}))
as the \celerite{} model \citep{foreman-mackey_fast_2017}
but it additionally takes into account a \leaf{} component.
We call \leaf{} matrix a general class of "close to diagonal" matrices which
encompasses banded, block-diagonal, and staircase matrices (see Eq.~(\ref{eq:leaf}) and Fig.~\ref{fig:leaf}).
For instance, the \leaf{} component of our model is well suited to account for calibration noise in radial velocity time series.

We illustrate the use of the \spleaf{} model
in the context of radial velocity time series
but the model is more general and could be adapted to other fields.
We reanalyze the HARPS radial velocity time series of \object{HD~136352}
using different noise models (see Sect.~\ref{sec:rvsystem})
and observe that the periodograms and FAP levels strongly depend on the chosen noise model.
We find that neglecting the short term correlated noise (short period quasiperiodic noise or calibration noise)
can lead to spurious detections of signals (underestimation of the FAP),
or to a poor detection power (over estimation of the FAP).
We thus show that the calibration noise,
which can be included in the \spleaf{} model,
has a substantial effect on detections.

\begin{acknowledgements}
  We thank the anonymous referee for their useful comments.
  We thank X.~Dumusque and C.~Lovis for fruitful discussions,
  and V.~Bourrier for finding the name \leaf{} while advocating against the use of \spleaf{}.
  We acknowledge financial support from the Swiss National Science Foundation (SNSF).
  This work has, in part, been carried out within the framework of
  the National Centre for Competence in Research PlanetS
  supported by SNSF.
\end{acknowledgements}

\bibliographystyle{aa}
\bibliography{spleaf}

\appendix

\section{Cholesky decomposition and solving in the preconditioned case}
\label{sec:overflowdetails}

In this appendix, we show how to adapt the algorithms of the Cholesky decomposition (Sect.~\ref{sec:choleskyspleaf})
and solving (Sect.~\ref{sec:detsolve}) to the preconditioned case.
As explained in Sect.~\ref{sec:overflow} \citep[and following][]{foreman-mackey_fast_2017},
we introduce the $(n-1)\times r$ preconditioning matrix $\phi$, and the preconditioned
matrices $\tilde{U}$ and $\tilde{V}$, such that:
\begin{equation}
  U_{i,s} V_{j,s} = \tilde{U}_{i,s} \tilde{V}_{j,s} \prod_{k=j}^{i-1} \phi_{k,s}.
\end{equation}
To stay consistent with this preconditioning, we additionally define
$\tilde{W}$, $\tilde{S}$, and $\tilde{Z}$ such that:
\begin{align}
  & U_{i,s} W_{j,s} = \tilde{U}_{i,s} \tilde{W}_{j,s} \prod_{k=j}^{i-1} \phi_{k,s},\nonumber \\
  & U_{i,s} S_{i,s,t} U_{i,t} = \tilde{U}_{i,s} \tilde{S}_{i,s,t} \tilde{U}_{i,t},\nonumber  \\
  & U_{j,s} Z_{i,j,s} = \tilde{U}_{j,s} \tilde{Z}_{i,j,s}.
\end{align}
The recursion formulas for the Cholesky decomposition in the preconditioned case
(see Eqs.~(\ref{eq:recS})-(\ref{eq:recW})) are:
\begin{align}
  \label{eq:recpreS}
  & \tilde{S}_{0,s,t} = 0,\nonumber                                                                                                        \\
  & \tilde{S}_{i,s,t} = \phi_{i-1,s}\phi_{i-1,t}\left(\tilde{S}_{i-1,s,t} + \tilde{W}_{i-1,s} D_{i-1} \tilde{W}_{i-1,t}\right)\quad (i>0), \\
  \label{eq:recpreZ}
  & \tilde{Z}_{i,i-b_i,s} = 0,\nonumber                                                                                                    \\
  & \tilde{Z}_{i,j,s} = \phi_{j-1,s}\left(\tilde{Z}_{i,j-1,s} + G_{i,j-1} D_{j-1} \tilde{W}_{j-1,s}\right)\quad (j>i-b_i),                 \\
  \label{eq:recpreG}
  & G_{i,j} = \frac{1}{D_j}\left(F_{i,j} - \sum_{k=\max(i-b_i, j-b_j)}^{j-1}\hspace{-6.5mm} G_{i,k} G_{j,k} D_k
  - \sum_{s}\tilde{U}_{j,s} \tilde{Z}_{i,j,s}\right),                                                                                       \\
  \label{eq:recpreD}
  & D_i = A_i - \sum_{s} \tilde{U}_{i,s} \left(\sum_{t} \tilde{S}_{i,s,t} \tilde{U}_{i,t} + 2 \tilde{Z}_{i,i,s}\right)
  - \sum_{k=i-b_i}^{i-1}\hspace{-1mm} G_{i,k}^2 D_k,                                                                                        \\
  \label{eq:recpreW}
  & \tilde{W}_{i,s} = \frac{1}{D_i}\left(\tilde{V}_{i,s} - \sum_{t} \tilde{S}_{i,s,t} \tilde{U}_{i,t} - \tilde{Z}_{i,i,s}\right).
\end{align}

The recursion formulas for the solving ($x = L^{-1}y$) in the preconditioned case
(see Eqs.~(\ref{eq:solveLf}) and (\ref{eq:solveLx})) are:
\begin{align}
  \label{eq:solveLpref}
  \tilde{f}_{0,s} & = 0,\nonumber                                                                        \\
  \tilde{f}_{i,s} & = \phi_{i-1,s}\left(\tilde{f}_{i-1,s} + \tilde{W}_{i-1,s} x_{i-1}\right)\quad (i>0), \\
  \label{eq:solveLprex}
  x_i             & = y_i - \sum_s \tilde{U}_{i,s} \tilde{f}_{i,s} - \sum_{j=i-b_i}^{i-1} G_{i,j} x_j,
\end{align}
where $\tilde{f}$ is defined such that
\begin{equation}
  U_{i,s} f_{i,s} = \tilde{U}_{i,s} \tilde{f}_{i,s}.
\end{equation}
The case of the dot product is very similar to the above (see Eqs.~(\ref{eq:dotLf}) and (\ref{eq:dotLy}))
as well as the dot product and solving with $L\t$.

\section{Backpropagation of the gradient for the \spleaf{} model}
\label{sec:backpropdetails}

In this section, we explain how to obtain gradient backpropagation algorithms for the \spleaf{} model.
\citet{foreman-mackey_scalable_2018} provided backpropagation algorithms
for the Cholesky decomposition and solving in the case of semiseparable matrices ($F=0$ in our notations).
We generalize this method to \spleaf{} matrices.
We do not detail here the full algorithms
but we rather describe the method used to obtain them
and refer the reader to the reference implementation (\spleafURL) for further details.

Let us first recall the steps required to evaluate the log-likelihood (see Sect.~\ref{sec:generalcase}):
\begin{itemize}
  \item Compute the deterministic part of the model $m(\theta)$, and the residuals $r=y-m(\theta)$;
  \item Compute the \spleaf{} representation of the covariance matrix $A(\alpha)$,
        $\tilde{U}(\alpha)$, $\tilde{V}(\alpha)$, $\phi(\alpha)$, $F(\alpha)$;
  \item Compute the Cholesky decomposition of the covariance matrix $D$, $\tilde{W}$, $G$;
  \item Compute the log-determinant $\displaystyle\ln\det(C) = \sum_i \ln D_i$;
  \item Solve for $u = L^{-1} r$;
  \item Compute $\displaystyle\chi^2 = u\t D^{-1} u = \sum_i \frac{u_i^2}{D_i}$;
  \item Compute $\ln\L = -\frac{1}{2} \left(\chi^2 + \ln\det(C) + n\ln\det(2\pi)\right)$.
\end{itemize}
We then need to compute the derivatives $\displaystyle\frac{\partial\ln\L}{\partial \theta}$
and $\displaystyle\frac{\partial\ln\L}{\partial \alpha}$.
There are typically two ways to achieve this, the forward and backward propagation of the gradient.
In the forward approach, computing the gradient (or the Jacobian matrix)
of $y_n(x) = f_n\circ\dots\circ f_2\circ f_1(x)$ is performed by first computing
$\nabla f_1(x)$ and propagating it using the relation,
\begin{equation}
  \label{eq:forwardprop}
  \nabla y_{k+1}(x) = \nabla f_{k+1}(y_{k}(x)) \nabla y_{k}(x),
\end{equation}
for $k=1\dots n-1$.
In the backward approach, we first compute $\nabla f_n(y_{n-1}(x))$, and propagate it using the relation:
\begin{equation}
  \label{eq:backprop}
  \nabla g_k(y_{k}(x)) = \nabla g_{k+1}(y_{k+1}(x)) \nabla f_{k+1}(y_k(x)),
\end{equation}
for $k=n-1\dots 1$, with $g_k = f_n\circ\dots\circ f_{k+1}$.
In both methods, we need to compute the gradient of each function appearing in the composition (each $f_i$).
The relative efficiency of both methods depends on the number of dimension of the parameter space and of the output space.
Let us note $p$ the number of parameters, and $m_k$ the number of dimension of $y_k(x) = f_k\circ\dots\circ f_2\circ f_1(x)$.
In the forward approach, each step consists in the computation of a $m_k \times p$ matrix as the dot product of
a $m_k \times m_{k-1}$ and a $m_{k-1} \times p$ matrices.
In the backward approach, each step consists of computing a $m_n \times m_k$ matrix as the dot product of
a $m_n \times m_{k+1}$ and a $m_{k+1} \times m_k$ matrices (with $m_0=p$).
Therefore, in the case $p<m_n$, the forward method should be more efficient,
while in the case $m_n<p$, the backward method should be faster.

In the case of the log-likelihood, we have $m_n = 1$ (the log-likelihood is a scalar function),
and the backward propagation should be preferred.
The backpropagation method to compute the gradient of the log-likelihood can be decomposed in the following
steps:
\begin{itemize}
  \item Compute $\displaystyle\frac{\partial\ln\L}{\partial u_i} = -\frac{u_i}{D_i}$;
  \item Compute $\displaystyle\frac{\partial\ln\L}{\partial D_i} = \frac{1}{2}\left(\left(\frac{u_i}{D_i}\right)^2 - \frac{1}{D_i}\right)$;
  \item Compute the gradient of $\ln\L$ with respect to $\tilde{U}$, $\tilde{W}$, $\phi$, $G$, and $r$
        by using a backpropagation algorithm for the solving ($u = L^{-1}r$),
        and the values of $\displaystyle\frac{\partial\ln\L}{\partial u_i}$;
  \item Use a backpropagation algorithm for the Cholesky decomposition to compute the gradient of $\ln\L$ with respect to
        $A$, $\tilde{U}$, $\tilde{V}$, $\phi$, and $F$;
  \item Backpropagate the gradient of $\ln\L$ with respect to the residuals to compute
        $\displaystyle\frac{\partial\ln\L}{\partial \theta} = \displaystyle\frac{\partial\ln\L}{\partial r}\displaystyle\frac{\partial r}{\partial \theta}$;
  \item Backpropagate the gradient of $\ln\L$ with respect to the \spleaf{} decomposition of the covariance to compute
        $\displaystyle\frac{\partial\ln\L}{\partial \alpha}$.
\end{itemize}

The $k$-th line of code appearing in the implementation of an algorithm (Cholesky decomposition, dot product $y=Lx$, solving, etc.)
can be seen as a function $f_k$, while the full code is the composition $y_n(x) = f_n\circ\dots\circ f_2\circ f_1(x)$.
In the case of the Cholesky decomposition, the vector $x$ represents all the entries of $A$, $\tilde{U}$, $\tilde{V}$, $\phi$, and $F$,
while the output $y_n(x)$ represents $D$, $\tilde{U}$, $\tilde{W}$, $\phi$, and $G$.
The backpropagation of the gradient for the Cholesky decomposition consists in computing the derivatives
$\displaystyle\frac{\partial h}{\partial A_i}$, etc., from the values of $\displaystyle\frac{\partial h}{\partial D_i}$, etc.,
for some function $h$.
Applying the backpropagation method described above (Eq.~(\ref{eq:backprop})) is equivalent
to reading the code of the algorithm in the reverse order
(starting from the last line, and reversing the order of each loop)
and backpropagating the gradient for each line.

Special care should be taken to ensure the stability of the method.
For instance, divisions by zero (or small numbers) should be avoided.
The only divisions that appear in the computation of the log-likelihood are the divisions by $D_i$
(in the Cholesky decomposition and in the computation of the $\chi^2$)
which are unavoidable but not problematic for a well conditioned matrix.
In the backpropagation algorithms, we also avoid any division other than divisions by $D_i$.
Let us illustrate why this is preferable with the update formula for the tensor $\tilde{S}$
involved in the Cholesky decomposition algorithm
(see Eq.~(\ref{eq:recpreS})).
As mentioned in Sect.~\ref{sec:choleskyspleaf},
when computing the Cholesky decomposition of a \spleaf{} matrix,
the $n\times r\times r$ tensor $\tilde{S}$ could be stored in memory as a much smaller $r\times r$ matrix
and updated in place using Eq.~(\ref{eq:recpreS}).
Then the final value of this $r\times r$ matrix $\tilde{S}_{n-1}$
could be used as an initial value in the backpropagation algorithm
and updated in place by computing $\tilde{S}_{i-1}$ from $\tilde{S}_{i}$ (see Eq.~(\ref{eq:recpreS})),
\begin{equation}
  \label{eq:invrecpreS}
  \tilde{S}_{i-1,s,t} = \frac{\tilde{S}_{i,s,t}}{\phi_{i-1,s}\phi_{i-1,t}} - \tilde{W}_{i-1,s} D_{i-1} \tilde{W}_{i-1,t}.
\end{equation}
This is done in the \celerite{} code \citep{foreman-mackey_scalable_2018} as it has a smaller memory footprint.
However, looking at the update formula~(\ref{eq:invrecpreS}), we can see that when $\phi_{i-1,s}\phi_{i-1,t} \approx 0$,
this turns out to be unstable numerically.
This issue could thus induce a wrong determination of the gradient,
which could slow down or prevent the convergence of minimization algorithms.
We thus store the full $\tilde{S}$ tensor in the Cholesky decomposition algorithm,
which increases the memory footprint of the algorithm but improves the efficiency and stability
of the backpropagation method.

\end{document}